\documentclass[amsmath,amssymb,twocolumn]{revtex4-2}

\usepackage{bm}
\usepackage{amsmath} 
\usepackage[usenames,dvipsnames]{color}
\usepackage{graphicx} 
\usepackage{amsfonts}
\usepackage{graphicx}
\usepackage{bm}
\usepackage{relsize}
\usepackage{braket}
\usepackage{bbold}
\usepackage{float}
\usepackage[unicode=true,colorlinks=true,urlcolor=blue,citecolor=blue]{hyperref}
\usepackage{hyperref}

\newcommand{\nix}[1]{}

\begin{document}

\title{Tunable resonant $s-p$ mixing of excitons in van der Waals heterostructures}

\author{Jiayu David Cao}
\email{jiayucao@buffalo.edu}
\thanks{Present address: University of Central Florida,
Orlando, Florida, USA.}

\author{Konstantin S.~Denisov}

\author{Igor \v{Z}uti\'{c}}
\email{zigor@buffalo.edu}

\affiliation{Department of Physics, University at Buffalo, State University of New York, Buffalo, New York 14260, USA} 

\begin{abstract}
Excitonic states of tightly-bound electron-hole pairs dominate 
the 
optical response in a growing class of two-dimensional (2D) materials and their van der Waals (vdW) heterostructures.  In transition metal dichalcogenides (TMDs) a useful guidance for the excitonic spectrum is 
the analogy with the states in the 2D hydrogen atom. From our symmetry analysis and solving the Bethe-Salpeter equations we find a much richer picture for excitons and predict their tunable resonant $s-p$ mixing.
The resonance is attained when the subband splitting matches the energy difference between the $1s$ and $2p_+$ (or $2p_-$)
excitons, resulting in the anticrossing of the spectral lines in the absorption as a function of the subband splitting. By focusing on TMDs modified by magnetic proximity, and gated 3R-stacked bilayer TMD, we corroborate the feasibility of such tunable spin splitting. The resulting tunable and bright $s-p$ excitons provide unexplored opportunities for their manipulation and enable optical detection of Rashba or interlayer coupling in vdW heterostructures.

\end{abstract}	

\date{\today}

\maketitle

\newpage

Two-dimensional (2D) materials with their suppressed screening and enhanced Coulomb interactions provide a versatile platform to study excitons with large binding energies~\cite{Wang2018:RMP,Huang2022:NN,Syperek2022:NC,Mak2016:NP,Unuchek2018:N,Scharf2019:PRB,Grzeszczyk2023:AM,Mittenzwey2025:PRL,Shao2025:NM,Galvani2016:PRB}.
The interplay between the Coulomb interaction, multiple valleys, and intrinsic spin-orbit coupling (SOC) in 2D materials influences the excitonic states and the resulting optical response~\cite{Xiao2012:PRL}. 
The sensitivity of excitonic properties to the neighboring regions in their heterostructures offers an important probe for proximitized materials~\cite{Zutic2019:MT}. The transformation of excitons by magnetic, SOC,
or charge-density wave proximity effects, as well as by the change in the band topology, or in twisted multilayers~\cite{Scharf2017:PRL,Joshi2022:APLM,Norden2019:NC,Serati2023:NL,Zhong2020:NN,Li2023:NM,Xu2020:PRL,Faria2023:NM,Volmer2023:npj,Wang2024:PRB}, then becomes a valuable fingerprint for emergent phenomena~\cite{Yang2024:PRB,VanTaun2017:PRX,Cao2024:PRB}.

\begin{figure}
    \centering
    \includegraphics[width=0.99\linewidth]{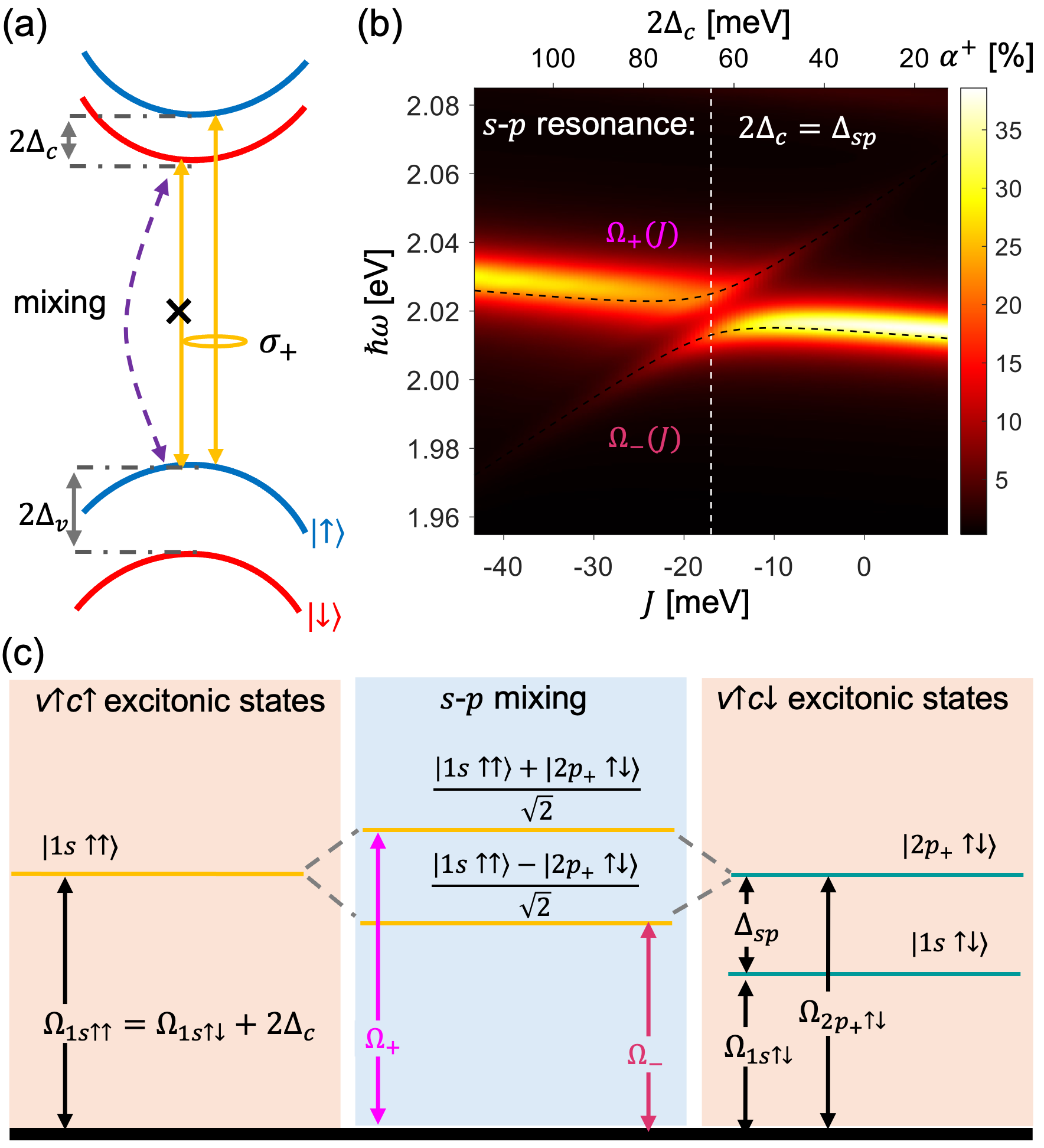}
    \caption{(a)
    Spin-forbidden and spin-allowed optical transitions in the $K$ valley of proximitized 
    ML TMDs, with valence ($v$) and conduction ($c$) band splitting, 
    $2\Delta_v \gg 2\Delta_c$. The mixing is introduced by a broken mirror symmetry. (b)~The evolution of the optical absorption and exciton energies, $\Omega_\pm$, as $2\Delta_c$ is tuned 
    by the proximity-induced exchange coupling 
    $J$, obtained from the Bethe-Salpeter equation. (c) 
    Exciton energy levels before and after ``turning on" of the mixing. $\Delta_{sp}=\Omega_{2p_+\uparrow\downarrow}-\Omega_{1s_+\uparrow\downarrow}$.
    Yellow (green) lines: 
    bright (dark) excitons.  The $s-p$ resonance condition $\Omega_{\uparrow\uparrow1s}=\Omega _{\downarrow\uparrow2p_+}$ or, equivalently, $2\Delta_c=\Delta_{sp}$ is demonstrated.}
    \label{fig:sp1}
    \vspace{-0.5cm}
\end{figure}

Transition-metal dichalcogenides (TMDs)~\cite{Wang2018:RMP}, one of the most studied 2D materials,  
offer a useful guidance for the excitonic spectrum from a 2D hydrogen atom, classified by the principal and angular quantum numbers.
For a simple isotropic spectrum, the Rydberg-like series, $1s$, $2s$, $2p,\dots$, and the $s$-excitons are usually active under single-photon processes. 
Beyond the simple isotropic two-band model,  
by adding other discrete degrees of freedom, such as valley or subband (spin or layer), 
the excitonic wave function acquires 
admixtures of different angular states resulting,
for example, in a partial brightening of $p$-like excitons~\cite{Cao2024:PRB,Hanan2015:PRB,Chirolli2020:PRB,Glazov2017:PRB,Zhu2023:PRL,Engdahl2024:arXiv}.

In this work we develop a 
theory of the resonant $s-p$ mixing of excitons and an enhanced brightening of $2p$ states in van der Waals (vdW) heterostructures with tunable subband splittings.
We illustrate the idea in Fig.~\ref{fig:sp1}
for a monolayer~(ML) TMD and excitons formed by spin-up~($\uparrow$) valance band states~($A$ excitons) in the $K$ valley. 
From Rashba SOC there is a mixing between $|1s{\uparrow\uparrow}\rangle$ and $|2p_+{\uparrow\downarrow}\rangle$ excitons, 
the first (second) spin is for electron from
valence ($v$) [conduction ($c$)] band, due to the subband spin mixing in the $c$ band. 
By including magnetic proximity effect, we change the 
$c$ band  splitting to push the energy of 
$|2p_+{\uparrow\downarrow}\rangle$ 
exciton close to that of $|1s{\uparrow\uparrow}\rangle$ 
[Fig.~\ref{fig:sp1}(c)]. We achieve a resonant $s-p$ mixing of excitons, a regime not accessible in intrinsic 2D insulators. 
The resulting anticrossing $s-p$ exciton doublet~[Fig.~\ref{fig:sp1}(b)] 
enhances the  brightening of $p$ states, instead of making them weakly active 
due to remote band admixtures~\cite{Cao2024:PRB,Hanan2015:PRB,Chirolli2020:PRB,Glazov2017:PRB, Zhang2022:PRL,Zhu2023:PRL}.

Our analysis uses 
the massive Dirac electron model 
with tunable subband splitting, 
complemented with solving
the the Bethe-Salpeter equation~(BSE)~\cite{Bechstedt:2016,Scharf2019:JPCM}, 
for magnetically-proximitized TMDs and for nonmagnetic bilayer (BL) TMDs with the gate-voltage controlled top (t) bottom (b) layer splitting. 
For the two coupled massive Dirac electrons the Hamiltonian  is
 \begin{equation}
       H
=\begin{pmatrix}
    \frac{E_g}{2} + {\Delta_{c}} & \hbar v_F k_- & 0 & 0\\
    \hbar v_F k_+ & -\frac{E_g}{2} + \Delta_{v}& \gamma^*& 0 \\
    0 & \gamma &\frac{E_g}{2} -\Delta_{c} & \hbar v_F k_-\\
    0 & 0 & \hbar v_F k_+ & -\frac{E_g}{2}  -\Delta_{v}
\end{pmatrix},
 \label{eq:H} 
\end{equation}
where $\Delta_{c,v}$ describe the subband splittings in $c$, $v$ bands, $E_g$ is the energy gap, and $v_F$ 
the Fermi velocity. 
The intersubband coupling due to broken out-of-plane mirror-symmetry, 
in the the lowest order of $k$,
is given by a parameter $\gamma$. 
The basis of $H$ is $|{c}{\uparrow}\rangle$, $|{v}{\uparrow}\rangle$, $|{c}{\downarrow}\rangle$, $|{v}{\downarrow}\rangle$. These spins correspond to the real spin 
for ML TMDs, or to the t/b-layer index for 
3R-stacked BL TMDs.
 
The interband transitions in $H$ are induced by $\sigma_+$ polarized photons, 
the time-reversal version of $H$ is used for $K'$-valley which couples to $\sigma_-$, with the spin allowed (forbidden) transitions optically active (inactive).
To demonstrate the appearance of the $s-p$ mixing, we use 
the effective mass approximation (EMA) in the limit of a large band gap compared with the subband splittings.
$H$ from Eq.~(\ref{eq:H}) will then be used only in BSE. 
Applying the downfolding procedure~\cite{Winkler:2003} in the second-order with respect to $c$-$v$ coupling, $\hbar v_F k /E_g$, we arrive at the effective $2\times 2$ Hamiltonian for $c$ and $v$ bands
\begin{equation}
    H_{c(v)} = \pm \left( \frac{E_g}{2} + \frac{\hbar^2 k^2}{2m^*} \right) + \Delta_{c(v)} \sigma_z 
    \pm (\alpha k_- \sigma_+ + {\rm c.c.} ),
    \label{eq:HcHv}
\end{equation}
where ${\bm{\sigma}}$ 
is the vector of Pauli matrices for subbands, 
$\sigma_\pm=(\sigma_x\pm i\sigma_y)/2$, $m^* = E_g/2v_F^2$ is the effective mass, and $\alpha= \gamma \hbar v_F/E_g$ is an effective subbands coupling strength.  
After downfolding the last term depends linearly on $k$, unlike $k$-independent $\gamma$ in $H$. 
$E_g$
and $\Delta_{c(v)}$ in Eq.~(\ref{eq:HcHv}) are renormalized by a small~$\gamma^2/E_g$.

The  exciton wave function  in the state $|S\rangle$ is written as  a vector: 
$\vec{\Psi}^S(\bm{r})=[\psi^S_{\uparrow\uparrow}
(\bm{r}),\psi^S_{\uparrow\downarrow}(\bm{r}),\psi^S_{\downarrow\uparrow}(\bm{r}),\psi^S_{\downarrow\downarrow}(\bm{r})]^T$, in the basis $| s_v s_c \rangle$, where the $s_{v(c)}$ run over subband indices ($\uparrow$, $\downarrow$) of the $v$($c$) bands, respectively, and $\bm{r}$ is the electron-hole (e-h) relative motion coordinate. 
The Schr\"{o}dinger equation for $\vec{\Psi}^S(\bm{r})$ is 
\begin{equation}
	[{I} \otimes H_c(\bm{r}) - H^*_v(\bm{r}) \otimes I + V(r)]\vec{\Psi}^S(\bm{r})=\Omega_S \vec{\Psi}^S(\bm{r}),
    \label{eq:Hr}
\end{equation}
where $I$ is the $2\times2$ unit matrix, $H_{c,v}(\bm{r})$
is obtained by replacing $k_{x,y}\to -i\partial_{x,y}$ in Eq.~(\ref{eq:HcHv}),
$V(r)$ models the attractive screened Coulomb interaction, 
and $\Omega_S$ is the exciton energy.
The matrix element of the $\sigma_+$ polarized optical transition from a ground state $|0\rangle$ to $|S\rangle$ described by $\vec{\Psi}^S(\bm{r})$ is given by
~\cite{Elliott1957:PR,SM} 
\begin{equation}
    \langle S|\sigma_+|0\rangle = v_F[\psi^S_{\uparrow\uparrow}(\bm{r}\to 0)+\psi^S_{\downarrow\downarrow}(\bm{r}\to0)],
    \label{eq:Elliott}
 \end{equation}
valid also for $\alpha \neq 0$. 
This expression 
(i) is nonzero for spin-allowed components of $\vec{\Psi}^S$ and  
(ii) accounts for the probability of finding electron and hole with the same spin at the same point in space ($\bm{r}\to0$), which is nonzero only for $s$-like states. 
The overall probability of the optical transition is determined by $|\psi_{\uparrow \uparrow} + \psi_{\downarrow\downarrow}|^2$, rather than by the sum of separate probabilities, $|\psi_{\uparrow \uparrow}|^2 + |\psi_{\downarrow\downarrow}|^2$.

At $\alpha=0$, 
$H_{c,v}$ 
in Eq.~(\ref{eq:Hr}) becomes diagonal, allowing us to classify excitons by 
$s_{v(c)}$, the principal $n$ and angular quantum numbers $m$~($1s$, $2s, 2p_{\pm}$...)~\cite{Yang1991:PRA,Chao1991:PRB}, 
the corresponding energies and wave functions, 
are given by
\begin{equation}
   \begin{split}
\Omega^0_{n,m;s_v s_c} &= E_g - s_{vz}\Delta_v + s_{cz} \Delta_c + E_{n,m},\\
\Vec{\Psi}^0_{n,m;s_v s_c}(\bm{r})&=  e^{i m \phi} \mathcal{R}_{n,m}(r)|s_v, s_c \rangle, 
\end{split} 
\label{eq:Rydberg}
\end{equation}
where $\bm{r} = (r, \phi)$,  
$\mathcal{R}_{n,m}(r)$ is the radial part of the wave function,  $E_{n,m}$ is the binding energy, and $s_{c,vz} = \pm 1$.
Both $\mathcal{R}_{n,m}$ and $E_{n,m}$ depend on $V(r)$: For $V(r) \propto 1/r^2$, 
we have the 2D Rydberg series $E_{n,m}=-{R_y}/{(n+1/2)^2}$.  
A more accurate Rytova-Keldysh potential~\cite{Rytova1967:PMP,Keldysh1979:JETP} gives 
$m$-dependent $E_{n,m}$.
For $s$ states ($m=0$),  
$\mathcal{R}_{n,0}(r\to 0) \neq 0$. Other excitons 
($m \neq 0$) have $\mathcal{R}_{n,m\neq 0}(r\to 0) = 0$.
Therefore, the 
matrix element from Eq.~(\ref{eq:Elliott})
 can be nonzero only for $s$ excitons, suggesting that $2p$ excitons are dark. 
 $\mathcal{R}_{1s}(r\to0)\gg\mathcal{R}_{2s}(r\to0)$ and 
 the 2D absorption spectrum 
 is dominated by the $1s$ excitons.

With $\alpha \neq 0$ due to the broken mirror symmetry, 
($n,m, s_{v}, s_c$) are no longer good quantum numbers: 
$H$ from Eq.~(\ref{eq:Hr})
commutes only with the operator of the total angular momentum defined as
\begin{equation}
\hat{J}_z=
    - i \partial_\phi
    -(1/2) I\otimes \sigma_z
    + (1/2)  \sigma_z
    \otimes I,
\label{eq:H0(r)}
\end{equation}
and a proper classification of excitonic states is based on the eigenvalues of $\hat{J}_z$,  
denoted by integer $l$.
We assume a weak intersubband coupling 
compared to the energy gap and the Coulomb interaction.
Treating $\alpha$ as a perturbation, the exciton wave function 
$\Vec{\Psi}_{l}(\bm{r})$ can be expressed 
in terms of unperturbed states
$\Vec{\Psi}^0_{n,m;s_v,s_c}$ 
from Eq.~(\ref{eq:Rydberg}) 
\begin{equation}
	\Vec{\Psi}_{l}(\bm{r})=
\begin{pmatrix}
	c_{\uparrow\uparrow} \mathcal{R}_{n_{\uparrow\uparrow},l}(r) \\
	c_{\uparrow\downarrow} \mathcal{R}_{n_{\uparrow\downarrow},l+1}(r) e^{i \phi} \\
	c_{\downarrow\uparrow} \mathcal{R}_{n_{\downarrow\uparrow},l-1}(r) e^{-i \phi} \\
	c_{\downarrow\downarrow} \mathcal{R}_{n_{\downarrow\downarrow},l}(r) 
\end{pmatrix}
e^{i l\phi},
\label{eq:Phi(r)}
\end{equation}
where $n_{s_vs_c}$  
is the principal quantum number of the $|s_vs_c\rangle$ state, and the coefficients $c_{s_vs_c}$ should be determined from the secular equation.  
$\Vec{\Psi}_{l}(\bm{r})$ 
displays the mixing  
between states with different $n,m$ and subband indices.  

While the considered mixing effects are valid for all $l \neq 0$, consisting predominantly of dark excitons, we focus on optically bright exciton configurations with $l=0$.
The $s-p$ mixing is realized at $l=0$ 
with the most efficient brightening of spin-forbidden $2p$ excitons. The mixing strength is inversely proportional to energy difference between the spin forbidden $2p$ state and spin allowed $1s$ states.
For example, the exciton in the state $|2p_{+} {\uparrow \downarrow} \rangle$ with 
$\Vec{\Psi}^0_{2p_{+}{\uparrow\downarrow}} = \mathcal{R}_{2p_{+}}(r) e^{ i \phi}|{\uparrow \downarrow}\rangle$ gets an admixture, $\delta \vec{\Psi}_{2p_{+}{\uparrow\downarrow}}$, 
consisting of $|1s {\uparrow\uparrow} \rangle$ and $|1s{\downarrow\downarrow} \rangle$
\begin{equation}
 \delta\vec{ \Psi}_{2p_+} = - i \alpha I_{sp} \left( \frac{ \vec{\Psi}_{1s \uparrow\uparrow}^0 }{\Delta_{sp} -2 \Delta_c} 
    +  \frac{ \vec{\Psi}_{1s \downarrow\downarrow}^0 }{\Delta_{sp} - 2\Delta_v}\right),
     \label{eq:spweak}
\end{equation}
where $I_{sp}= 2\pi\int rdr \mathcal{R}_{1s}(r)[\partial_r \mathcal{R}_{2p}(r)+\mathcal{R}_{2p}(r)/r]$, 
while $\Delta_{sp}$ is the energy difference between $1s$ and $2p$ excitons with the same spin configuration before ``turning on" the mixing [see also Fig.~\ref{fig:sp1}(c)] and Eq.~(\ref{eq:Rydberg}) is used to calculate exciton energy differences. The
$s-p$ mixing yields 
a partial brightening of $2p_+$ exciton:
\noindent{the} matrix element from Eq.~(\ref{eq:Elliott}) scales as $\propto \alpha v_F I_{sp} \mathcal{R}_{1s}(0)$ at a fixed 
$\Delta_{c(v)}$. 
With a tunable $\Delta_{c(v)}$,
we consider 
$|2\Delta_{c(v)}|=\Delta_{sp}$ for 
a resonant condition of $\Omega_{2p_+ {\uparrow\downarrow}}^0=\Omega^0_{1s \uparrow\uparrow}~\mathrm{or}~\Omega_{1s \downarrow\downarrow}^0$.  
Since the denominators in Eq.~(\ref{eq:spweak}) diverge, 
one has to solve a  
full secular equation for $c_{s_vs_c}$. 

To confirm these predictions and accurately 
include the Coulomb interaction in the absorption
spectra,  we use the BSE where the state $|S\rangle$ exciton
is expressed 
as a many-body wave function 
built from the eigenstates of the single-particle  
$H$ from Eq.~(\ref{eq:H}), as $|S\rangle=\sum_{vc\bm{k}}\mathcal{A}^S_{vc}(\bm{k})|vc\bm{k}\rangle$, with $|vc\bm{k}\rangle=\hat{c}^\dagger_{c\bm{k}}\hat{c}_{v\bm{k}}|0\rangle$, where the ground state, $|0\rangle$, consists of fully occupied $v$ bands and empty $c$ bands,
$c^\dagger_{c\bm{k}}$ ($c_{v\bm{k}}$) is the creation (annihilation) of an electron in 
the eigenstate 
$|c_\pm,\bm{k}\rangle, |v_\pm,\bm{k}\rangle$ of $H$,  
where $c_\pm, v_\pm$ are subband
indices. 
The exciton envelope wave function
in $k$-space, $\mathcal{A}_{vc}^S(\bm{k})$, and the corresponding exciton energy, $\Omega^S$, are calculated 
by solving the 
BSE~\cite{SM}

\begin{align}
        \Omega_{S}&
	\mathcal{A}^S_{vc}(\bm{k}) =\left[\epsilon_{c}(\bm{k})-\epsilon_{v}(\bm{k})\right]\mathcal{A}_{vc}^S(\bm{k})\nonumber \\
    &-\frac{1}{A}\sum_{v'c'\bm{k}'}V(|\bm{k}-\bm{k}'|)\langle c \bm{k}|c' \bm{k}'\rangle\langle v' \bm{k}'|v \bm{k}\rangle \mathcal{A}_{v'c'}^S(\bm{k}'),  
    \label{eq:BSE}
\end{align}
where $\epsilon_{c,v}(\bm{k})$ are the single particle energies of $H$, 
$A$ is the area of our system,
$V(q)=2\pi e^2/(\varepsilon q+r_0q^2)$ is the Fourier transform of the screened Coulomb interaction described by the Rytova-Keldysh potential, with $\varepsilon$ for the background dielectric constant and $r_0$ for the screening radius, see~\cite{SM}.

Using 
$\vec{\mathcal{A}}^S(\bm{k})=[\mathcal{A}^S_{++}(\bm{k}),\mathcal{A}^S_{+-}(\bm{k}),\mathcal{A}^S_{-+}(\bm{k}),\mathcal{A}^S_{--}(\bm{k})]^T$, 
we can 
relate  
$\Vec{\mathcal{A}}^S(\bm{k})$
to $\vec{\Psi}_l(\bm{r})$ from Eq.~(\ref{eq:Phi(r)})
by performing the Fourier and unitary transformations (changing 
the basis from $|s_v, s_c \rangle$ to $|vc \bm{k} \rangle$), 
allowing us to write
\begin{equation}    \label{eq:A(k)_l}
    \vec{\mathcal{A}}_l(\bm{k})=\begin{pmatrix}
        \mathcal{Q}_{++}(k)\\
        \mathcal{Q}_{+-}(k)e^{i\phi_k}\\
        \mathcal{Q}_{-+}(k)e^{-i\phi_k}\\
        \mathcal{Q}_{--}(k)
    \end{pmatrix}
    e^{il\phi_k},
\end{equation}
with $\bm{k} = (k, \phi_k)$, 
and the functions $\mathcal{Q}_{vc}$ can be expressed through $\mathcal{R}_{n,m}$, see~\cite{SM}. 
Generally, $\mathcal{Q}_{vc}$ is an admixture of all $4$ components of $\vec{\Psi}_l$.
Only in the limit of $\alpha k\ll \Delta_{c,v}$, 
when the basis $|vc\bm{k} \rangle \approx|s_vs_c\rangle$,  
$\mathcal{Q}_{vc}$ becomes determined by the Fourier transform
of the single component $\mathcal{R}_{nm}$. 
 For each component of $\vec{\mathcal{A}}^S(\bm{k})$, we can define the winding number, $w$, which counts the number of times the phase in $\mathcal{A}^S_{vc}(\bm{k})$ rotates when completing a full circle in $\phi_k$~\cite{Zhang2018:PRL,Cao2018:PRL}. 
 $w=l,l+1,l-1,l$ for four components of $\vec{\mathcal{A}}_l$. We note that, $\mathcal{A}_{vc}^S(\bm{k})$ is gauge-dependent, a gauge consistent with our analytical EMA method is used~\cite{SM}.

The absorption coefficient of photon $\sigma_\pm$ at angular frequency $\omega$ is evaluated from
\begin{equation}
	\alpha^\pm(\omega)=\frac{4 e^2 \pi^2}{ c\omega}\sum_{S} \left|\langle S|\sigma_{\pm}|0\rangle\right|^2 \delta(\hbar\omega-\Omega^S),
\label{eq:abs}
\end{equation}
where $\langle S|\sigma_{\pm}|0\rangle=\sum_{vc\bm{k}}[\mathcal{A}^S_{vc}(\bm{k})]^*\mathcal{D}_{vc}(\bm{k})/A$.
The optical transition matrix element is $\mathcal{D}_{vc}^\pm(\bm{k}) =\langle c\bm{k}|\hat{v}_\pm|v\bm{k}\rangle$, 
with  
$\hat{v}_\pm=(\hat{v}_x\pm i\hat{v}_y)/\sqrt2$, and 
$\hat{v}_{x,y}=\partial H/ \partial(\hbar k_{x,y})$. 
To fit the absorption with experiments, we replace the $\delta$-function by a Lorentzian of a 
finite spectral width, 
$\Gamma = 5$~meV.

\begin{figure}[b]
    \centering
    \includegraphics[width=0.99\linewidth]{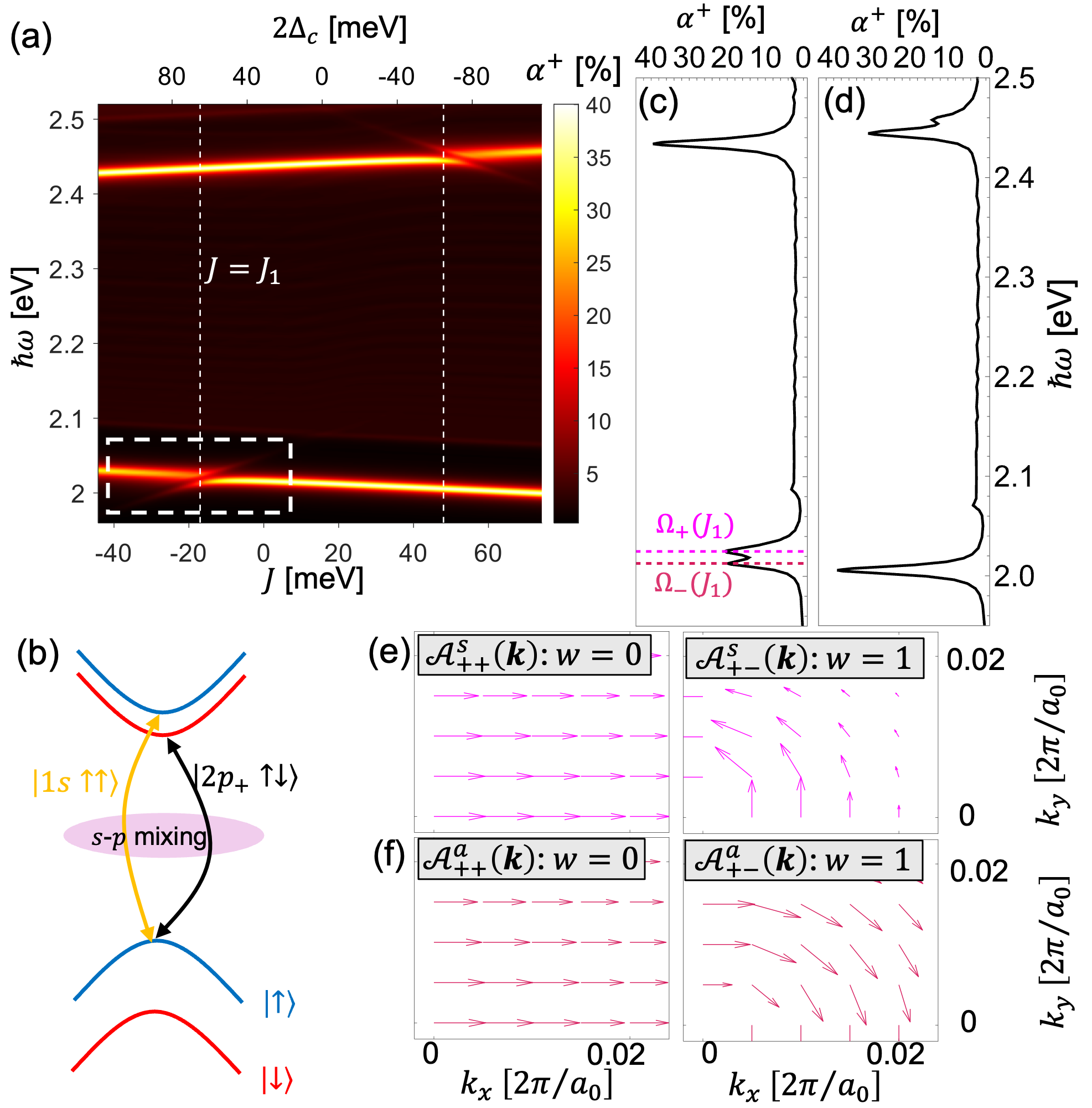}
    \caption{BSE results for ML TMD:
    (a)~
    Brightening of the $|2p_+{\uparrow\downarrow}\rangle$ and $|2p_-{\downarrow\uparrow}\rangle$~excitons with different 
    $J$. Anticrossings appear in bottom-left 
    [its enhanced view, white rectangle, in Fig.~\ref{fig:sp2}(b)]
    and top-right region. 
    (b)~Mixing of the $|1s{\uparrow\uparrow}\rangle$ and $|2p_+{\uparrow\downarrow}\rangle$ excitonic states. (c), 
     (d)~The absorption spectra at two $s-p$ resonances. Marked as vertical white dashed lines in (a). (e),  (f)~Phases of dominant components of 
    $\vec{\mathcal{A}}^{s,a}(\bm{k})$, see Eq.~(\ref{eq:A_ML}).
    The direction (length) of an arrow denotes the phase and (the magnitude).  
    Exciton energies at the resonance and their
    evolution with $J$ are marked in (c) and Fig.~\ref{fig:sp1}(b).
    }
    \label{fig:sp2}
\end{figure}

We first consider magnetically proximitized ML TMDs. We use an out-of-plane orientation of the substrate magnetization, with spin-subband splittings: $\Delta_{c(v)}=\lambda_{c(v)} - J_{c(v)}$, 
where $\lambda_{c(v)}$ and
$J_{c(v)}$ are the intrinsic SOC and the exchange splitting due to the magnetic proximity effect in $c$ ($v$) 
bands of ML TMDs.
Typically, $\lambda_{v}$ of a few hundred 
meV exceeds greatly $\lambda_c, J_{c}$.
$\Delta_v \gg \Delta_c$ implies the formation of exciton series $A$ and $B$ related to spin states $v=|\uparrow\rangle$ and $|\downarrow \rangle$ and separated by the energy $\Delta_v$. 
This allows us to consider the $s-p$ mixing within the $A$- and $B$-series independently, neglecting the $A$-$B$ coupling effects [the first two and the last two components are considered independently in Eq.~(\ref{eq:Phi(r)})]. 
The resonant condition for the $s-p$ mixing is then realized by varying $\Delta_c$ through $J_c$. 
The substrate-induced breaking of the mirror symmetry also leads to the Rashba SOC,  
providing  the spin-subband coupling 
needed for the $s-p$ mixing. The Rashba SOC takes the form $\gamma \to i\gamma_R$ in Eq.~(\ref{eq:H}), corresponding to $\pm \alpha (k_x \sigma_y - k_y \sigma_x)$ in $H_{c(v)}$, Eq.~(\ref{eq:HcHv}), with $\alpha\to i\alpha_R$. The SOC-strength $\alpha_R \sim 0.15$~eV\AA{}, consistent with typical Rashba splitting in 2D~\cite{Stier2023:PRB,Soumyanarayanan2016:N,Durnev2014:PRB}.

In Fig.~\ref{fig:sp2} we show the 
calculated absorption as the function of frequency and 
the proximity-induced exchange interaction, $J$, with $J_c=J$, $J_v=0.76J$, guided 
by $\mathrm{WSe}_2$/EuS parameters~\cite{SM}.
The absorption is dominated by two resonant lines of $A$:$1s$ ($\hbar \omega \approx 2.02 $~eV) and $B$:$1s$ ($\hbar \omega \approx 2.43$~eV) excitonic transitions, showing a linear slope $\pm0.24$ with $J$. 
The anticrossings from the $s-p$ resonances are 
at $2 \Delta_c = \pm\Delta_{sp}=\pm65\;$meV for $A$ or $B$ 
exciton series. The resonant absorption spectrum has 
a double-peak frequency profile,  
shown in Figs.~\ref{fig:sp2}(c),~\ref{fig:sp2}(d).

To further clarify the obtained resonant $s-p$ mixing, we focus on 
$A$-excitons (negative $J_c$, see Fig.~\ref{fig:sp1}), and ignore a weak admixture of $B$-exciton states. With the basis 
$|1s{\uparrow\uparrow\rangle}$ and $|2p_+{\uparrow\downarrow}\rangle$, the effective Hamiltonian is
\begin{equation}\label{eq:HML}
    H_{\mathrm{ML}} = 
    \begin{pmatrix}
    \Omega^{0}_{\uparrow\uparrow 1s} & \alpha_R I_{sp} \\
\alpha_R I_{sp} & \Omega^{0}_{\uparrow\downarrow 2p_+} 
    \end{pmatrix}, 
\end{equation}
giving us the
energies $\Omega_{\pm}=\bar{\Omega}\pm \sqrt{(\Delta_{\Omega })^2+(\alpha_RI_{sp})^2}$, where $\bar{\Omega}=[\Omega^0_{\uparrow\uparrow1s}+\Omega^0_{\uparrow\downarrow2p_+}]/2$, and the detuning is $\Delta_{\Omega }=(\Omega^0_{\uparrow\uparrow1s}-\Omega^0_{\uparrow\downarrow2p_+})/2$. 
The energies, $\Omega_\pm$, shown in Fig.~\ref{fig:sp1}(b) with dashed black lines as a function of $J$ or $2\Delta_c$, 
feature the anticrossing and are in agreement with the exact absorption obtained from BSE in Fig.~\ref{fig:sp1}(b).

At the resonance, $\Delta_{\Omega}=0$, we have 
$\Omega_\pm=\bar{\Omega}\pm \alpha_R I_{sp}$, 
with the energy splitting, $\delta \Omega = 2\alpha_R I_{sp}$, proportional to $\alpha_R$ and the eigenstates given by the symmetric (s) and antisymmetric (a) combinations of $|1s \uparrow \uparrow \rangle$ and $|2p_+ \uparrow \downarrow\rangle$ 
\begin{equation}    \label{eq:spML}
    \vec{\Psi}_{s,a} = 
    \left( \vec{\Psi}_{1s {\uparrow \uparrow}}^0(r)  \pm \vec{\Psi}_{2p_+ {\uparrow \downarrow}}^0 \right)/\sqrt{2}. 
\end{equation}
We obtain the numerical estimate, $\delta \Omega=\Omega_+(J_1)-\Omega_-(J_1) \approx 16$~meV [see Fig.~\ref{fig:sp2}(c)], defining the  beating frequency between the states s/a.
Here, $J_1$ is the magnitude of $J$ realizing
the exact $s$-$p$ resonance.
The oscillator strength becomes equally distributed between two peaks separated by $\delta \Omega$, shown in Figs.~\ref{fig:sp2}(a), \ref{fig:sp2}(c). 
To observe the double peak structure in absorption, as in Fig.~\ref{fig:sp2}(c), one needs to cool sample to suppress the exciton broadening, $\Gamma \lesssim \delta \Omega$. 

It is instructive to discuss the exciton eigenstates Eq.~(\ref{eq:spML}) in the BSE form 
\begin{equation}\label{eq:A_ML}
\vec{\mathcal{A}}^{s,a}(\bm{k}) = [\mathcal{Q}^{s,a}_{++}(k) , \mathcal{Q}_{+-}^{s,a}(k)e^{i\phi_k},0,0]^T. 
\end{equation}
The winding number structure of $\vec{\mathcal{A}}^{s,a}(\bm{k})$ is 
confirmed from BSE, as shown in 
Figs.~\ref{fig:sp2}(e), \ref{fig:sp2}(f), with the winding numbers $w=0, 1$. 
The last two components are zero, since contributions from $B$-excitons are negligibly small.
The resonant character of the $s-p$ mixing 
is not explicitly accessible in $\mathcal{Q}^{s,a}$, which complicates
the analysis of this effect in 2D materials only from the BSE method. 
The actual $k$-dependence of $\mathcal{Q}(k)$ does not display signatures of s/a structures from Eq.~(\ref{eq:spML}) 
due to a large spread of the 2D exciton wave function 
in $k$-space with $\alpha k_{ex} \gtrsim \Delta_c$, 
where $k_{ex}$ is the $k$-space exciton Bohr radius. 

\begin{figure}[h]
    \centering
    \includegraphics[width=0.99\linewidth]{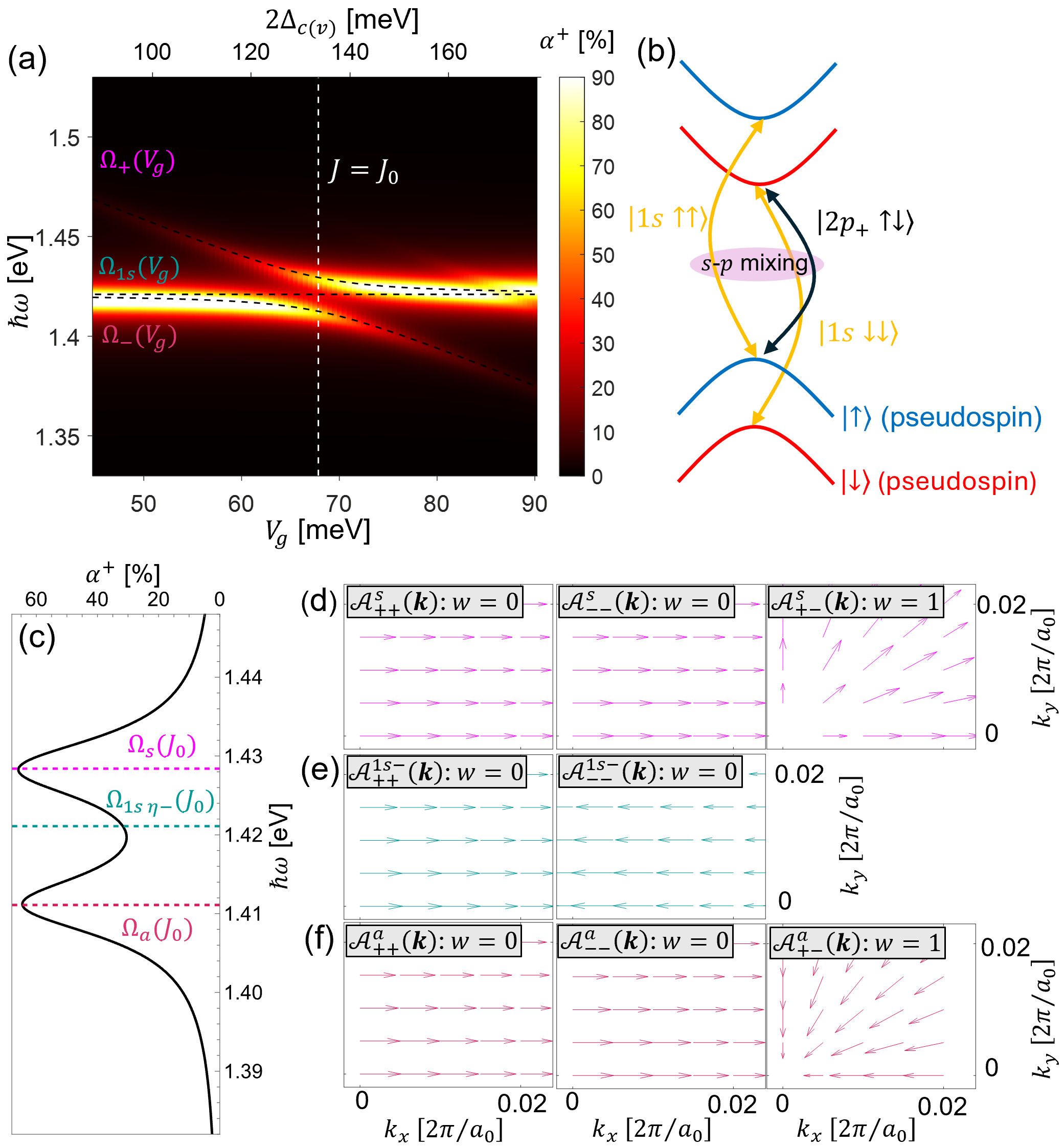}
    \caption{BSE results for 3R stacked BL TMD:
    (a)~Evolution of the absorption as the subband splitting $2\Delta_{c(v)}$ is tuned through out-of-plane gate voltage $V_g$.
    Here, $J_0$ is the magnitude of $J$ realizing the exact $s-p$ resonance.
    (b)~Mixing of the intralayer $1s$ excitons and the interlayer 
    $|2p_+{\uparrow\downarrow}\rangle$ exciton. 
    The $|{\uparrow}({\downarrow)}\rangle$ pseudospin denotes the t(b) layer index.  (c)~The absorption at the resonance, as marked by the white dashed line in (a). (d)-(f).~Phases of the 
    dominant components of $\vec{\mathcal{A}}^{s,a}(\bm{k})$ and $\vec{\mathcal{A}}^{1s-}(\bm{k})$. Corresponding exciton energies at the resonance and evolution with $V_g$ are marked in (c) and (a).}
    \label{fig:sp3}
\end{figure}

We next consider $H$ which 
retains the electron-hole and the subband symmetry 
($\Delta_c=\Delta_v$). This 
can be realized in a spinless model of 3R-stacked 
BL TMD,
where only the lower(upper) $c$($v$) bands of the t and b layers are taken into account.  Here our spinor notation $|{\uparrow}( {\downarrow}) \rangle$ denotes t(b) layer states,
shown in Fig.~\ref{fig:sp3}.
The symmetric subband splittings are 
controlled by an out-of-plane gate voltage $V_g$, following $2\Delta_{c(v)}= V_g$. In our model, while an applied electric is responsible for tunable $\Delta_{c(v)}$, we do not consider higher order effects, such as the Stark effect, 
where the spectral weight of the main excitonic peaks will slightly decrease as the field is increased~\cite{Scharf2016:PRB}. The mixing is realized due to an interlayer coupling described by real $\gamma$ in $H$,  
which gives the Dirac term 
$\pm\alpha_{I} (k_x \sigma_x + k_y \sigma_y)$, with $\alpha\to\alpha_I$ in the 
downfolded $H_{c(v)}$ from Eq.~(\ref{eq:HcHv}).

In this case, the exciton energies in the 
$|{\uparrow\uparrow}\rangle$ and 
$|{\downarrow\downarrow}\rangle$ 
series are the same, suggesting the use of the basis 
$|\eta_\pm\rangle =(|{\uparrow\uparrow}\rangle \pm 
|{\downarrow\downarrow}\rangle)/\sqrt{2}$. To identify the structure of the resonant $s-p$ mixing, we write the effective Hamiltonian in the basis  $|1s\,\eta_-\rangle$, $|1s\,\eta_+\rangle$ and 
$|2p_+{\uparrow\downarrow}\rangle$ 
\begin{equation}
    H_{BL}=
    \begin{pmatrix}
        \Omega^0_{1s\,\eta_-} & 0 & 0\\
        0 & \Omega_{1s\,\eta_+}^0 & -i\sqrt2\alpha_I I_{sp}\\
        0 & i\sqrt2\alpha_I I_{sp} &\Omega^0_{\uparrow\downarrow2p_+}
    \end{pmatrix},
    \label{eq:HBL}
\end{equation}
where $\Omega^0_{1s\,\eta_\pm}=\Omega^0_{1s\uparrow\uparrow}=\Omega^0_{1s\downarrow\downarrow}$. 
The lower block of Eq.~(\ref{eq:HBL}) has the same form as Eq.~(\ref{eq:HML}). 
As a consequence, 
the absorption spectrum, the exciton energies evolution, and the winding numbers 
of $\vec{\mathcal{A}}^{s,a}(\bm{k})$ 
are the same as 
in the previous case in Fig.~\ref{fig:sp2}. 
Unlike in ML TMDs, we realize a dark spin-allowed $1s$ exciton: $|1s\,\eta_-\rangle$. 
This exciton configuration consists of optically bright $1s$ states but becomes 
completely dark because of its wave function 
anti-symmetrization, as in Eq.~(\ref{eq:Elliott}).
The corresponding wave function anti-symmetrization is confirmed in Fig.~\ref{fig:sp3}(e). 
The constant energy of this exciton is a straight horizontal line in Fig.~\ref{fig:sp3}(a) which lies between the two bright $s-p$ mixed states.
Despite an effective mixing between 3 exciton states, the calculated absorption displays only a single anticrossing feature accompanied by the formation of a double peak structure in its frequency profile, as shown in Fig.~\ref{fig:sp3}(c), making the picture qualitatively similar to the magnetically-proximitized TMDs with Rashba SOC in Fig.~\ref{fig:sp1}(b) and Fig.~\ref{fig:sp2}(a).

Our predictions have focused on the two specific examples of magnetically-proximitized monolayer TMDs and 3R-stacked bilayer TMDs. However, the principle of resonant mixing of excitonic states
with different orbital symmetry can be extended to other materials systems, and have many further implications. 
With a  small twist angle in (nearly)commensurate bilayer TMDs an applied gate voltage may also brighten interlayer $R^M_h$ $2p$ excitons, where the effective Hamiltonian can be written in the form of $H_{BL}$ in Eq.~(\ref{eq:HBL})~\cite{Wu2018:PRB,Huang2022:NN,Regan2022:NRM}.
It would be instructive to consider 
resonant mixing in the recently found valley-dependent magnetic proximity effects~\cite{Choi2023:NM}, leading to an intriguing
possibility of the topological effects and gap closing at only one ($K$ or $K'$) valley~\cite{Zhou2023:NM} and the resulting change
of the helicity of the emitted light.
It is interesting to study resonance mixing of excitons in monolayer TMDs under arbitrarily oriented external magnetic fields,
where the excitonic mixing is recognized in such systems by GW-BSE method~\cite{deFariaJr2025:arXiv}.

With the 
importance of Rashba spin-orbit coupling to the mechanism 
of spin-mixing, recently demonstrated to also dominate the spin relaxation of
interlayer MoSe$_2$/MoS$_2$ excitons~\cite{Mittenzwey2025:PRL}, it would be helpful examine Janus TMDs~\cite{Gan2022:AM}
and generalize spin-orbit coupling beyond the usual $k$-linear dependence. 
There is a growing class of materials where its $k$-cubic dependence is significant 
and can alter the orbital symmetry of the proximity effects~\cite{Alidoust2021:PRB},
while $s-p$ mixing could be replaced by other orbital symmetry mixing. 
For these studies, our employed approach
which combines the solutions of Bethe-Salpeter equation
with the transparent symmetry analysis and massive Dirac
electron model could also provide useful insights.
Moreover, instead of a weak mixing regime of Ref.~\cite{Glazov2017:PRB}, 
we expect that the resonant mixing regime can become more prominent in the two-photon processes.

While the presence of $s-p$ and mixing of other symmetries
is more generic, the resonant character of the mixing
and the associated anticrossing features have distinct and largely unexplored properties. This resonant mixing 
becomes possible since with a larger binding energy of
excitons, due to the reduced screening of Coulomb interaction, there is also a larger 1$s$ to 2$p$ energy separation, close to the magnitude of the spin splitting in the conduction band. A challenge for future materials studies is to identify systems with an even larger 
and tunable spin splitting to match these two energies. 
From magnetic proximity effects,
such experimentally realized spin splitting in TMDs is
$\sim 20\;$meV~\cite{Zutic2019:MT,Norden2019:NC,Wang2022:npjCM},
already much larger than what is feasible with the strongest
applied magnetic field.
Taken together, 
the excitonic signatures and resonant spin mixing in systems with tunable spin splitting
can offer a sensitive probe for various forms of spin-orbit coupling as well design
optical response in spintronic devices not limited to 
magnetoresistance~\cite{Lee2014:APL,Liang2017:NC,Tornatzky2021:APL,Lindemann2019:N,Dainone2024:N}.

\textit{Acknowledgements.}
We thank G. Xu, B. Scharf, and H. Dery, for valuable discussions.
This work was supported by the U.S. Department
of Energy, Office of Science, Basic Energy Sciences under Award 
No. DE-SC0004890. Computational resources were
provided by the UB Center for Computational Research.

\bibliography{spX}
\end{document}